# Solar Energy Generation in Three Dimensions

Marco Bernardi[1], Nicola Ferralis[1], Jin H. Wan[2], Rachelle Villalon[3], & Jeffrey C. Grossman[1,†]

[1] Department of Materials Science and Engineering,

[2] Department of Mathematics,

[3] Department of Architecture,

Massachusetts Institute of Technology, 77 Massachusetts Avenue, Cambridge, MA 02139-4307

[†] e-mail: jcg@mit.edu

We formulate, solve computationally and study experimentally the problem of collecting solar energy in three dimensions.[1–5] We demonstrate that absorbers and reflectors can be combined in the absence of sun tracking to build three-dimensional photovoltaic (3DPV) structures that can generate measured energy densities (energy per base area, kWh/m$^2$) higher by a factor of 2–20 than stationary flat PV panels, versus an increase by a factor of 1.3–1.8 achieved with a flat panel using dual-axis sun tracking.[6] The increased energy density is countered by a higher solar cell area per generated energy for 3DPV compared to flat panel design (by a factor of 1.5–4 in our conditions), but accompanied by a vast range of improvements. 3DPV structures are steadier sources of solar energy generation at all latitudes: they can double the number of peak power generation hours and dramatically reduce the seasonal, latitude and weather variations of solar energy generation compared to a flat panel design. Self-supporting 3D shapes can create new schemes for PV installation and the increased energy density can facilitate the use of cheaper thin film materials in area-limited applications. Our findings suggest that harnessing solar energy in three dimensions can open new avenues towards Terawatt-scale generation.

**MAIN TEXT**

Converting the abundant flow of solar power to the Earth (87 PW) into affordable electricity is an enormous challenge, limited only by human ingenuity. Photovoltaic (PV) conversion has emerged as a rapidly expanding technology capable of reaching GW-scale electric power generation with the highest power density among renewable sources of 20–40 W/m$^2$.[7,8] The main barriers to widespread adoption of PV technology include system costs (3–5 $/Watt-peak) of which ~60% is due to installation costs,[9] the limited number of peak insolation hours available in most locations (further reduced by weather), and the requirement of a minimum threshold power density for cheaper thin-film technologies to become feasible for residential or commercial rooftop installations.

The main approach applied so far to alleviate these problems has been to search for lower-cost active layers with higher power conversion efficiencies. However, efficiency improvements can only partially reduce the installation costs and cannot change the pattern of solar energy generation, since these aspects are related to the PV system design.

A commonly adopted design consists of flat panels arranged on a flat surface – often a rooftop imposing further geometrical constraints – that yields far-from-optimal coupling with the Sun's trajectory. Sun-tracking systems can extend the range of useful peak hours, but add significant costs and are not well suited for residential or commercial installations due to the use of expensive and bulky movable parts.

The flat design of PV systems contrasts with the three-dimensionality of sunlight collecting structures found in Nature.[3,4] Two main physical reasons underlying the advantages of collecting light in 3D are the presence of multiple orientations of the absorbers that allow for the effective capture of off-peak sunlight, and the re-absorption of light reflected within the 3D structure.

We recently employed computer simulations (Ref. 5) to show that 3D photovoltaic (3DPV) structures can increase the generated energy density (energy per footprint area, kWh/m$^2$) by a factor linear in the structure height, for a given day and location. Optimal shapes derived using a genetic algorithm approach include a cubic box open at the top and a cubic box with funnel-like shaped faces, both capable in principle of doubling the daily energy density.[5] The higher area of PV material per unit of generated energy compared to flat panel designs is a main disadvantage of 3DPV, although this is alleviated by the fact that the module is not the main cost in PV installations at present, and the PV outlay will become increasingly dominated by non-module costs in the near future.[9] Additional practical challenges include inexpensive 3D fabrication routes and optimization of the electrical connections between the cells to avoid power losses.

Despite the enormous potential of macroscopic 3DPV structures, the lack of a comprehensive optimization approach and systematic study of the benefits in different seasons, locations and weather conditions, combined with the fact that the module has until only recently dominated the total cost of PV, have thus far limited the advancement of 3DPV as a groundbreaking concept and technology.

Here, we demonstrate that 3DPV structures can be realized practically and can dramatically improve solar energy generation: compared to a flat panel, they can nearly double the number of peak hours available for solar energy generation, provide a measured increase in the energy density by a factor of ~2–20 without sun tracking with even higher figures in the case of cloudy weather, and reduce the large variability in solar energy generation with latitude and season found in non-tracking flat panels. 3DPV structures additionally enable the design of effective sunlight concentrators using fixed mirrors.

We establish and implement numerically a general formalism to calculate the energy generated over a period of time, at any location on Earth, by a 3D assembly of $N$ solar cells of arbitrary shape, orientation, conversion efficiency and optical properties (Supplementary Information). The calculations account for inter-cell shading, Air-Mass effects in the incident solar energy and angle-dependent reflection of unpolarized light.[10] The Sun's trajectory is computed for the particular day and location using an algorithm developed by Reda *et al.*[11,12] Weather is not explicitly taken into account in the simulations and unless otherwise stated all the simulated energy values in this work assume clear weather.

Once the 3DPV structure (here for convenience broken down into triangles) has been defined, the generated energy can be expressed as an objective function of the cell coordinates that can be maximized using standard Monte Carlo (MC) simulated annealing and genetic algorithm (GA) optimization techniques,[13–17] both implemented here. The two main forces operating during the maximization of energy generation in 3D are the avoidance of inter-cell shading and the re-absorption of light reflected by other cells, with an intricate trade-off (dependent on the Sun's trajectory) typical of complex systems.

While here the focus is on electricity generation, the general computational approach we have implemented could allow for the optimization of a wide range of human activities that rely on sunlight collection, including heating, food crops, wine-making, and sustainable buildings.

In order to study 3DPV systems experimentally, we fabricated[18] and tested simple 3DPV structures consisting of a cube open at the top covered by solar cells both on the interior and exterior surfaces (here referred to as an open cube structure), a similar open parallelepiped of the same base area but twice as high, and a tower with ridged faces (Fig. 1a, and Supplementary Information). The structures are made of, respectively, 9, 17 and 32 commercially available Si solar panels.

Next, we measured the performance of the 3DPV structures. A flat panel was tested indoors under simulated solar light for validation of our simulations at different tilt angles to the light source (Fig. 1b), while measurements for all 3DPV shapes in Fig. 1a were collected outdoors under direct sunlight illumination (Fig. 1c–e and Supplementary Information). We validated the calculations from our computer code by comparing with experimental results for identical conditions (Fig. 1b,c and Supplementary Information) and found excellent agreement between the two, thus confirming the reliability of our code.[19]

The measured performance of a design as simple as the open cube under direct sunlight illumination on a summer day (Jun 16$^{th}$) shows clearly the benefits of 3DPV compared to the conventional flat design (Fig. 1c): a near doubling in the daily energy generation of 2.25 Wh (2.27 Wh in the simulation) was measured compared to 1.22 Wh (1.01 Wh in the simulation) for a flat solar cell of the same base area under the same conditions, resulting from an increase in both the number of hours of peak power generation and the power output throughout the day. The number of hours over which power generation was approximately constant is more than doubled for the 3DPV case compared to the flat panel, and extends between 1 hr after sunrise and until 1 hr before sunset.

Larger gains over a flat panel can be achieved using taller and more complex structures such as the open parallelepiped and ridged tower (Fig. 1d), with increases during the winter season even further enhanced compared to the summer. For example, the daily energy generation measured in clear weather (Fig. 1d) for a winter day (Nov. 18[th]) expressed as a ratio to the energy generated by a flat panel tested under the same conditions was 4.88 for the open cube, 8.49 for the parallelepiped and 21.5 for the tower.

The excess solar cell area per unit generated energy used for the 3DPV structures compared to the flat panel case was in the range of 1.5–4 for the cases examined here, with a minimum value of 1.5 corresponding to the tower case in the winter and a maximum value of ~4 for the cube in the summer.

Taller and more complex structures show an increasingly inhomogeneous cell illumination pattern with a higher number of partially shaded cells (Supplementary Movie), an effect that can introduce power losses[20] and ultimately reduce the overall energy gain. We found that such power losses are mainly determined by the presence of parasitic dark currents in the shaded cells, and we were able to successfully minimize these losses with the addition of blocking diodes in series with each panel in the structure (Supplementary Information).

We used the same outdoor testing apparatus to measure the performance of 3DPV systems under different weather conditions during the same week as the clear weather results in Fig. 1d. Our data shows that the diffuse light induced by clouds, rain and mist can be captured much more efficiently in 3DPV systems compared to flat panels, leading to increased energy generation enhancement factors for cloudy weather compared to clear weather (Fig. 1d,e). The relative decrease in generated energy due to clouds is thus less significant for a 3D structure than for a flat panel and hence 3DPV systems are a source of renewable electricity less impacted by weather conditions.

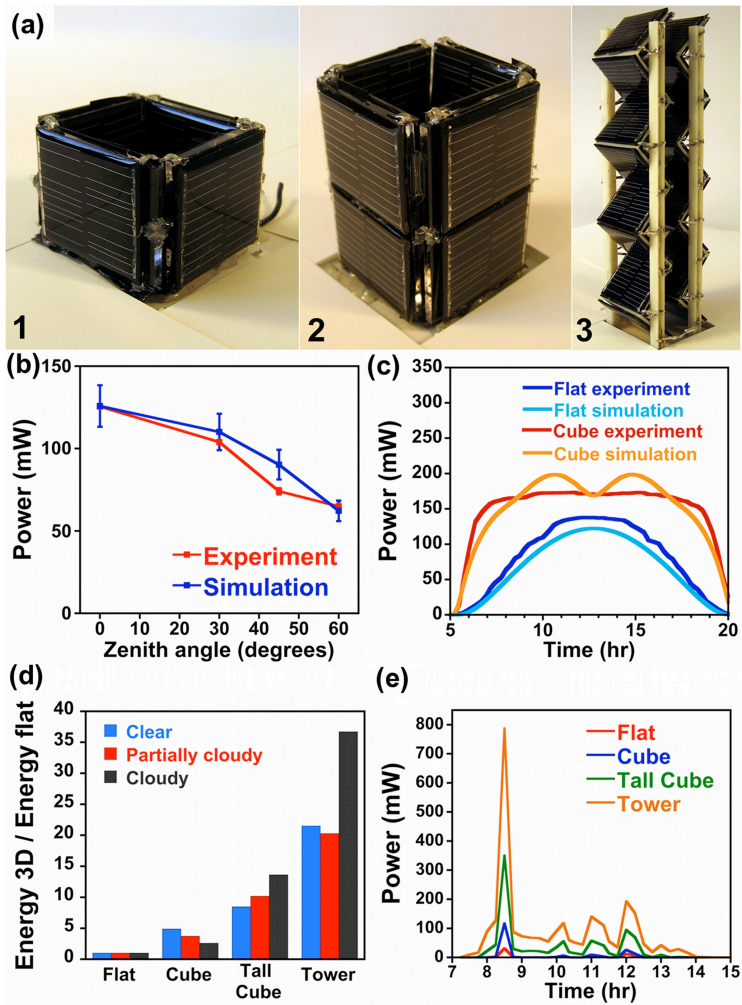

**Fig. 1** (a) 3DPV structures made using Si solar cells with area 3x3 cm$^2$. From left to right, an open cube (1), an open parallelepiped twice as tall (2), and a tower (3). The structures are made up, respectively, of 9, 17, and 32 solar cells. (b) Power generated by a flat Si panel at various tilt angles measured under simulated solar light illumination, and comparison with computer simulation. (c) Both measured and simulated power during a single sunny day for the flat panel and open cube, showing a maximal range of hours of constant power generation and an almost double energy density output for the 3DPV case compared to the flat panel. (d) Energy generated by the structures shown in (a) under different weather conditions, expressed as a ratio to the energy generated by a flat panel under the same weather conditions. Comparison of the black and blue bars for the case of the parallelepiped and tower shows how structures of higher aspect ratio than the open cube can further outperform a flat panel on a cloudy day compared to a clear day. The parallelepiped in (a) is referred here as "tall cube". (e) Power generated *vs.* time for the data of cloudy weather shown in (d).

In order to assess the effects of season and latitude on 3DPV performance, we studied the annual energy generation of 3DPV systems – a quantity strongly dependent on the coupling to the Sun's trajectory throughout the year – at different locations on Earth. We performed computer simulations of the energy generated by 3DPV structures over a full year at latitudes between 35° South to 65° North (almost all inhabited land), with an approximate latitude increase of 10° between locations and for over 20 cities in the world (Supplementary Information). These results are compared with data for fixed horizontal panels (from our simulations) and for both fixed flat panels with optimal orientation and using dual-axis sun tracking (from the literature, see Ref. 6).

Optimal static panel orientation can afford an increase in annual generated energy density (kWh/m$^2$ year) compared to a flat horizontal panel by a factor of 1.1–1.25.[6] Dual-axis tracking provides at present the best way to dynamically couple a PV panel to the Sun's trajectory, and can yield an increase of annual generated energy by a factor of 1.35–1.8 compared to a flat horizontal panel,[6] at the cost of using expensive movable parts to track the Sun's position.

For comparison, we calculated the same ratio (defined as *Y* here) of annual generated *energy density* for simple 3DPV structures to that of a flat horizontal panel of same base area, at several different latitudes (Fig. 2a). Even with a simple open cube structure, a large increase in the annual energy generation compared to a flat horizontal panel is found for 3DPV, with values of *Y* in the range 2.1–3.8, increasing monotonically from the equator to the poles. This trend compensates the lower ground insolation at larger latitudes to give an overall density of generated energy with significantly lower variation between locations at different latitudes for the 3DPV case compared to a flat panel (Table S2 in Supplementary Information).

When compared to flat panels with optimal orientation (Ref. 6, or from our calculations with similar results), an increase in the generated energy density in the range of 1.8–3 is found, thus still superior to the dual-axis tracking case.

For latitudes with maximal population density (between 50° N and 25° N)[21] values of *Y* are in the range of 2.5–3, suggesting that 3DPV structures can be used to increase the energy density (and consequently enable cheaper PV technologies) in geographical areas where future PV installations will abound.

The ratio of generated energy from a 3D structure to that of a flat panel increases from summer to winter (Fig. 2b) by a larger factor at higher latitudes, implying that 3DPV has lower variation in the energy generation due to season, for the same physical reason leading to reduced latitude variability – namely, a greater ability to collect sunlight when the sun is at low elevation angles compared to a flat panel.

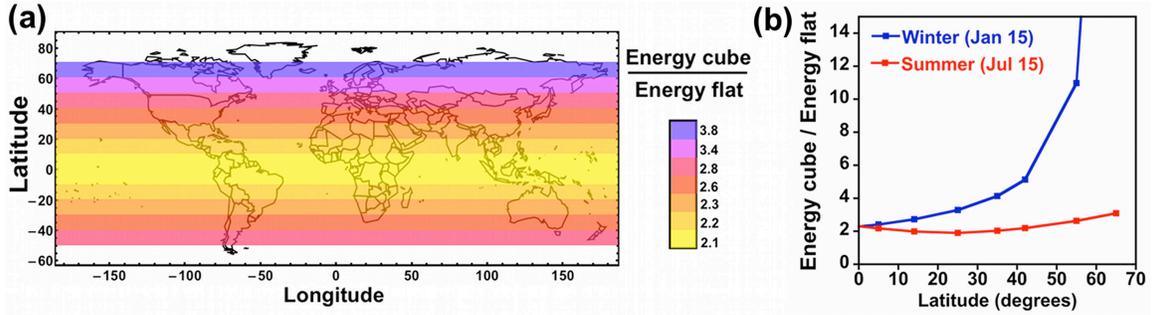

**Fig. 2** (a) Density plot of the variable *Y*, defined as the ratio of the annual energy density for an open cube 3DPV structure to that of a flat horizontal panel of same base area. Values of *Y* in the range 2.1–3.8 found here for static 3DPV structures largely exceed those predicted for dual-axis tracking. (b) Ratio of energy generated by an open cube compared to a flat panel for different season. 3DPV outperforms a flat panel by a larger amount during the winter and at higher latitudes due to the increased ability to use sunlight from lower elevations in the sky. The winter and summer labels refer to the Northern hemisphere; the curves would look the same for the Southern hemisphere provided the difference in season is taken into account.

Further possibilities to exploit solar energy generation in 3D include incorporating mirrors together with PV panels within the structure, with the aim of concentrating sunlight without sun-tracking systems, in contrast to existing concentrating technologies. Structures made of a combination of mirrors and solar panels were optimized using a simulated annealing optimization scheme. The concentration ratio (a figure of merit) is defined here as a ratio between the energy per unit area of active material generated with and without mirrors.[22]

A highest concentration of ~3.5 was obtained for maximal mirror area within a fixed simulation volume (Fig. 3a). The best concentrating structure consisted of a solar cell cutting the body diagonal of the simulation box and enclosed within two regions of mirrors in an "open flower" configuration facing the Sun (Fig. 3b). In this high-concentration limit, the use of a given amount of PV material is optimal for the 3DPV case: the energy per unit of PV active material is almost as high as for the flat panel case, yet with an energy generation 25% higher than the latter. On the other hand, a higher mirror area causes a decrease in the generated energy density, thus defining two opposite limits for volumetric solar energy generation (Fig. 3a): maximal energy per footprint area (3DPV case) and maximal energy per active material area (flat panel case). This further elucidates the difference between sunlight collection in two and three dimensions, and illustrates the extra design flexibility inherent to the use of 3D structures.

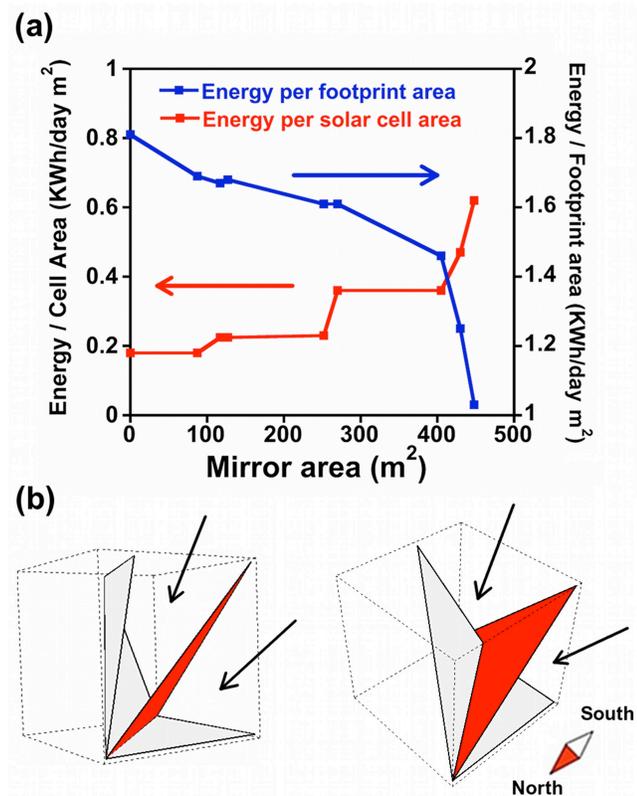

**Fig. 3** (a) Concentration of light by means of mirrors is quantified by the increase in the energy per unit solar cell area. For 3D solutions provided by the MC algorithm with a 10 m side cubic simulation box, the red curve describes the energy obtained in a day per unit area of solar cells. In the absence of mirrors, 3DPV optimizes the energy/footprint area (blue curve) rather than the energy per solar cell area. The latter can be optimized by sunlight concentration, as seen from the opposite trend of the two curves. A maximal concentration ratio of ~3.5 is inferred by comparing the values at the two ends of the red curve. (b) Best-concentrating configuration of mirrors (light gray) and solar cells (red) in a 10 m side cubic volume; a simplified structure extracted from the MC optimization is shown here. It consists of a solar panel arranged between two mirrors and resembles a flower open towards the sunlight direction (South in the figure). The black arrows show the direction of incident sunlight.

In addition to intriguing fundamental aspects, 3D solar collecting structures show tremendous promise for practical applications. Potential 3DPV technologies could include structures that ship flat and expand to fill a volume in an origami-like manner, for ground or flat-roof installation, or chargers for electric-powered vehicles in urban areas, or in sustainable buildings using novel semitransparent flexible PV cells incorporated in walls and windows. Two such cases are examined in detail in the Supplementary Information: a 3D electric bike charger prototype currently under development in our lab and a 50 m tall building with the surface completely coated with solar panels.

In closing, we observe that a comparative cost analysis between 3DPV and flat panel design is far from simple: apart from the higher number of panels used per unit energy in 3DPV, estimates of the installation costs and solar cell wiring costs are necessary, together with an estimate of the benefits of having a larger number of peak hours during the day. A detailed study would benefit from using the concept of levelized cost of energy,[23] although this is beyond the scope of the present work.

In summary, the striking range of improvements imparted by three-dimensionality to static solar collecting structures stems from their optimal coupling with the Sun's trajectory. 3DPV structures using simple shapes and electrical connections largely outperform flat panels of the same base area, and show promise for embedding PV systems in the urban environment beyond the flat panel form on rooftops. Computer design facilitates the prediction of generated energy and optimal shapes, and will be an indispensable tool for optimizing solar energy generation. Our results show that 3D sunlight collection has the potential to serve as a paradigm shift in solar energy conversion toward the Terawatt scale.

in the experiment. At the solar noon, the simulated power shows a marked dip for the open cube case, also present to a smaller extent in the experiment. Such difference can be attributed to the ideality of the simulated structure, where the infinitesimal cell thickness and the ideal due-South orientation completely cancel the contribution from side cells in the simulation at the solar noon, and also to electrical effects as discussed below in the paper. For these reasons, the very small discrepancy (1 %) found for the energy generated by the cube is a result of error compensation, a discrepancy between the simulation and the experiment of the order of 5–15 % is not uncommon even when the main physical effects have been captured in the simulation, as in the flat panel case in Fig. 1c.

20 A. Woyte, J. Nijs and R. Belmans, *Solar Energy,* 2003, **74**, 217.

21 W. Tobler, *Proceedings, AutoCarto 13*, Seattle (April 1997).

22 It must be noted that 3DPV does not normally optimize the energy per unit area of active material, but rather optimizes the energy generated from a given volume and hence for a given footprint area, as discussed below in the paper.

23 S. B. Darling, F. You, T. Veselka and A. Velosa, *Energy Environ. Sci.*, 2011, **4**, 3133-3139.

# Solar Energy Generation in Three Dimensions: Supplementary Information

Marco Bernardi, Nicola Ferralis, Jin H. Wan, Rachelle Villalon, and Jeffrey C. Grossman

The supplementary information is divided into two main sections, covering respectively:

1 – Methods and Materials

2 – Supplementary Text, further divided into:

    2.1 – 3DPV computer code, optimization, and validation

    2.2 – Details of all simulations

    2.3 – Power balancing and weather effects

    2.4 – Examples of applications of 3DPV

## 1. METHODS AND MATERIALS

The 3DPV structures were fabricated using commercially available Si solar cells purchased from Solarbotics (type SCC3733, 37x33mm, monocrystalline Si) with nominal open-circuit voltage and short-circuit current of 6.7 V and 20 mA, respectively, and AM1.5 efficiency of 10 % verified independently with a solar simulator. The Si active layer is protected by a 1 mm epoxy layer with refractive index $n = 1.6$ that yields an overall cell reflectivity at normal incidence $R \approx 14\ \%$. The panels were mounted onto a 3D-printed plastic frame for the particular structure of choice, and electrically connected in parallel through a main parallel bus, with blocking diodes placed in series with each cell. 3D-printed frames were realized using Fused Deposition Modeling (FDM) manufacturing technology with strict tolerances (0.005"), in a 3D printer purchased from Stratasys Inc. Frame models were realized in acrylonitrile butadiene styrene polymer starting from a 3D digital polygon mesh of the object. The same 3D digital files were converted into input files with proper format for the 3DPV computer code, thus allowing us to simulate 3DPV structures identical to those fabricated experimentally. This method is flexible and can be adapted to manufacture arbitrary shapes. The indoor characterization of the 3DPV structures was performed using a solar simulator with a 1300 W xenon arc lamp (Newport Corporation, model 91194) fitted with a global AM 1.5 filter and calibrated to 1000 W/m$^2$ intensity. In order to perform angle-dependent measurements of the current-voltage characteristics, solar panels were mounted on supports with predetermined angles to allow adjustment of the orientation with respect to the incoming light. Since commonly employed solar simulators are designed to operate at a very well defined distance between the light source and the tested flat panel, any variation of such distance has a significant effect on the illumination intensity and thus on the power output, which made outdoor testing necessary for accurate results for all 3D shapes. The outdoor solar cell characterization was performed in Cambridge on the roof of building 13 of the MIT campus during the months from June to December 2011. Up to four structures (including a flat panel used for reference) were tested simultaneously during each session to allow direct comparisons independent of weather conditions. The orientation of the structures was carefully checked with the use of a compass. The current-voltage characteristics were measured at time intervals of 12–15 min using a custom-made acquisition system.

## 2. SUPPLEMENTARY TEXT

### 2.1 – 3DPV COMPUTER CODE, OPTIMIZATION, AND VALIDATION

**Energy calculation.**

The main routine of our code performs computation of the total energy absorbed during any given period of time and at any given location on Earth by a 3D assembly of panels of given reflectivity and power conversion efficiency. This routine incorporates key differences compared to the one used in Ref. 5. For example, the use of Air Mass (AM) correction for solar flux allows for simulations during different seasons and at different latitudes, with reliable calculation of power curves and total energy. The computation has been generalized to account for cells of different efficiency and reflectivity within the structure, thus expanding the design opportunities to systems such as solar energy concentrators, where the mirrors are added as cells with zero efficiency and 100% reflectivity. The code has been extended to incorporate a start and an end date, so that simulations over any interval of time are possible. The Fresnel equations employed now assume unpolarized sunlight. All aspects of the code were carefully tested (see below), and new optimization methods were added to find energy maxima as discussed below.

Our algorithm considers a 3D assembly of $N$ panels of arbitrary efficiency and optical properties, where the $l^{th}$ panel has refractive index $n_l$ and conversion efficiency $\eta_l$ ($l = 1, 2, \ldots, N$, and $\eta_l = 0$ for mirrors). The energy $E$ (kWh/day) generated in a given day and location can be computed as:

$$E = \sum_{k=1}^{24/\Delta t} P_k \cdot \Delta t$$

where $\Delta t$ is a time-step (in hours) in the solar trajectory allowing for converged energy, and $P_k$ is the total power generated at the $k^{th}$ solar time-step. The total energy over a period of time is obtained by looping over the days composing the period, and summing the energies generated during each day.

The key quantity $P_k$ can be expanded perturbatively:

$$P_k = P_k^{(0)} + P_k^{(1)} + ... + P_k^{(m)} + ...$$

where the $m^{th}$ term accounts for $m$ reflections of a light ray that initially hit the structure – and is thus of order $R^m$, where $R$ is the average reflectivity of the absorbers – so that for most cases of practical interest, an expansion up to $m = 1$ suffices.

Explicitly, $P_k^{(0)}$ and $P_k^{(1)}$ can be written as:

$$P_k^{(0)} = \sum_{l=1}^{N} I_k \cdot \eta_l \cdot A_{l,\text{eff}}[1 - R_l(\theta_{l,k})]cos(\theta_{l,k}) \qquad (1)$$

$$P_k^{(1)} = \sum_{l=1}^{N} \{[I_k \cdot \eta_l \cdot A_{l,\text{eff}} R_l(\theta_{l,k})cos(\theta_{l,k})] \cdot \eta_s[1 - R_s(\alpha_{ls,k})]\} \qquad (2)$$

where $I_k$ is the incident energy flux from the Sun at the $k^{th}$ time-step (and includes a correction for Air-Mass effects), $A_{l,\text{eff}}$ is the unshaded area of the $l^{th}$ cell, and $R_l(\theta_{l,k})$ is the reflectivity of the $l^{th}$ cell for an incidence angle (from the local normal) $\theta_{l,k}$ at the $k^{th}$ time-step, that is calculated through the Fresnel equations for unpolarized incident light. In eq. (2) the residual power after the absorption of direct sunlight (first square bracket) is transmitted from the $l^{th}$ cell and redirected with specular reflection to the $s^{th}$ cell that gets first hit by the reflected ray. The $s^{th}$ cell absorbs it according to its efficiency $\eta_s$ and to its reflectivity calculated using the angle of incidence $\alpha_{ls,k}$ with the reflected ray. In practice, both formulas are computed by setting up a fine converged grid for each cell (normally $g$ =10,000 grid-points) so that all quantities are computed by looping over sub-cells of area equal to $A / g$ ($A$ is the area of a given triangular cell area), which also removes the need to summing $s$ over the subset of cells hit by the reflected light coming out of a given cell.

The empirical expression used to calculate the intensity of incident light on the Earth surface $I_k$ (W/m$^2$) with AM correction for the Sun's position at the $k^{th}$ time-step is:[24]

$$I_k \equiv I(\beta_k) = 1.1 \cdot 1353 \cdot 0.7^{\left[\frac{1}{\cos(\beta_k)}\right]^{0.678}} \qquad (3)$$

where $\beta_k$ is the Zenith angle of the sun ray with the Earth's local normal at the $k^{th}$ solar time-step, 1353 W/m$^2$ is the solar constant, the factor 1.1 takes into account (though in an elementary way) diffuse radiation, and the third factor contains the absorption from the atmosphere and an empirical AM correction. The angle $\beta_k$ is calculated at each step of the Sun's trajectory for the particular day and location using a solar vector obtained from the solar position algorithm developed by Reda et al.[11] and incorporated into our code.

Dispersion effects (dependence of optical properties on radiation wavelength) and weather conditions are not taken into account and are the main approximations of our model. Dispersion effects are fairly difficult to include, and would increase the computation time by a factor of 10–100. Weather effects require reliable weather information (*e.g.* from satellites), and seem interesting to explore in light of recent work on optimization of PV output based on weather by Lave et al.[25]

**Optimization algorithms.**

Our code uses genetic algorithm (GA) and simulated-annealing Monte Carlo (MC) methods to maximize the energy $E$ generated in a given day in the phase space constituted by the panels' coordinates. The GA algorithm was used as described in Ref. 5. Briefly, candidate 3D structures are combined using operations based on three principles of natural selection (selection, recombination, and mutation), using a GA algorithm adapted from Ref. 15. The "tournament without replacement" selection scheme was used,[26] in which $s$ structures from the current population are chosen randomly and the one of highest fitness proceeds to the mating pool, until a desired pool size is reached. In our simulations $s = 2$, and the fitness function corresponds to the energy $E$ produced in one day by the given structure.

The recombination step randomly combines 3D structures in the mating pool and with some probability (here 80%) crosses their triangle coordinates, causing the swapping of whole triangles. A two-point crossover recombination method was employed, wherein two indices are selected at random in the list of coordinates composing the chromosomes, and then the entire string of coordinates in between is traded between the pair of solutions.

Finally, the mutation operator slightly perturbs each coordinate, for the purpose of searching more efficiently the coordinates space. These three operations are performed until convergence is reached (usually 10,000–50,000 simulation steps), and a 3DPV structure with maximal energy production is achieved. The number of grid-points per triangular cell was fixed to 100 during most optimizations to limit computation time, and following the optimization optimal structures were re-examined using 10,000 grid-points as in all other simulations.

The MC algorithm was used to optimize structures of mixed optical properties, where we chose trial moves that preserve the optical properties of the single cells to favor the convergence of the optimization process. Our MC algorithm uses a standard Metropolis scheme for acceptance/rejection of trial moves, and a fictitious temperature $T$ that is decreased during the simulation according to a specified cooling schedule.[13] Trial moves consisted in the change of a randomly chosen set of coordinates of the candidate structure. A number of coordinates varying between 1 and $9N$ ($N$ is the number of triangles) were translated randomly within a cubic box of side 1–100 % the simulation box side (with periodic boundary conditions), thus determining a change $\Delta E$ in the total energy. When $\Delta E > 0$ the move is accepted, whereas when $\Delta E < 0$ the move is accepted with probability

$$P = e^{-|\Delta E|/T}.$$

When a move is accepted, the structure is updated and a new random move is performed. Most MC simulations consisted of 100,000 steps with a converged value of the final energy. The code implements both power law and inverse-log cooling schedules;[13] in most simulations we used the inverse-log cooling schedule

$$T(t) = \frac{c}{a + \log t}.$$

Average $\Delta E$ values for the given trial move were determined prior to running the optimization with a short simulation (1000 steps). Parameters for the cooling schedule were determined by imposing a temperature value such that the initial acceptance rate is $P=0.99$ and the final acceptance is $P=10^{-5}$, with a method detailed in Ref. 13.

The GA and MC algorithms gave consistent results for optimization of 1, 2, 3, 4, 10, 20, 50 cells in a 10x10x10 m$^3$ cubic simulation box (not shown here), suggesting that both algorithms are capable of finding energy values near the global maximum using less than 100,000 steps. Since the main cost of the simulation is the energy computation routine, the cost is comparable for both algorithms. For example, a 100,000 steps long simulation with 20 cells is completed in 1–2 days on a single processor. Parallelization of the code using standard MPI library is in the agenda, and could cut the computation time by a factor linear in the number of processors.

**Validation tests.**

The reliability of our code was checked using a large number of tests in a multitude of conditions, some of which are reported here and in the paper. Computation of the inter-cell shading and reflected energy was tested using a structure and a Sun trajectory shown in Fig. S1a. In a day and location where the apparent Sun trajectory goes from East to West keeping 90° (or 270°) azimuth angle at all times (*e.g.* Sept. 19 at latitude 1°, longitude -71°, and GMC time -5),[12] a tall wall (50 m height) was placed vertical to the ground, and a small square mirror (1 m side) was placed 10 m away from the wall and tilted so to completely reflect incident light to the middle height of the wall when the Sun is at the Zenith (11:35AM for our case). Until the Sun goes over the wall, the mirrors are shaded; they only start receiving light only at approx. 11AM, and keep reflecting light to the wall almost until the sunset. Comparison is made in Fig. S1b between the power generated at different times of the day for the two following cases:

1- the wall does not generate energy (its efficiency is set to zero), and the small mirror cells absorb and convert all incident light with unit efficiency (blue curve);

2- the wall absorbs all of the incident light, both reflected by the mirrors (here with zero conversion efficiency) and direct incident sunlight. For this case, difference in power between with and without mirrors is calculated, and represents the gain due to energy transferred by the mirrors and absorbed by the wall (red curve).

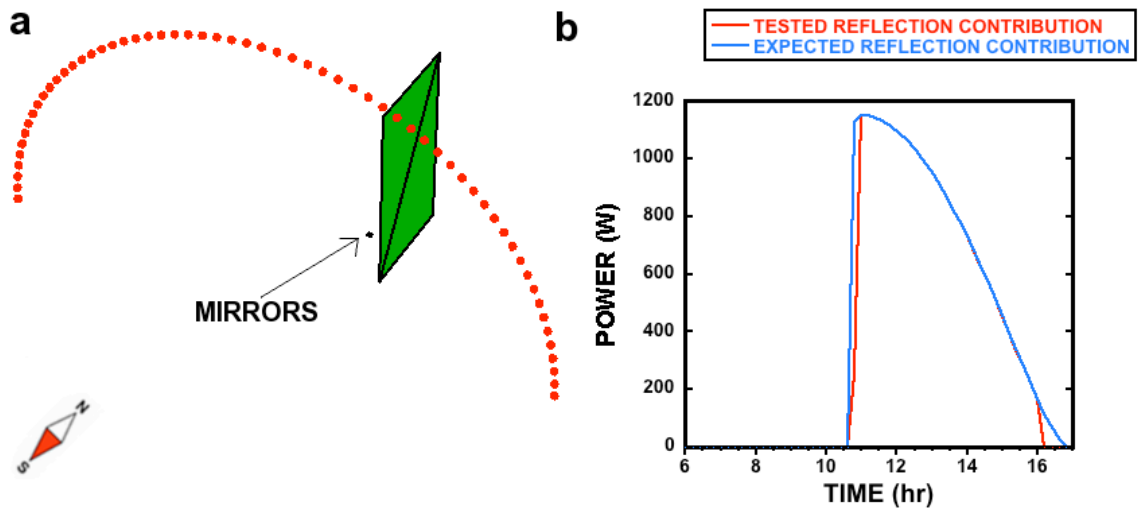

**Fig. S1** (a) Tested trajectory (red dots) re-scaled by a factor of 200,000. The wall is shown in green, and the mirrors indicated by the arrow. (b) Comparison of expected and calculated power contribution from the mirror, validating our reflection and shading algorithms.

Cases 1 and 2 should yield almost identical power values at all times between 11AM and sunset, since geometrical optics imposes that rays incident on the mirrors are reflected completely to the wall in this situation, and the power incident on the mirrors (expected contribution, blue curve in Fig. S1b) is transferred and absorbed completely by the wall. Besides, shading of the small mirrors is expected to occur almost until 11AM. The perfect matching observed in Fig. S1b between the predicted and observed curve in the useful time interval demonstrates at once that the code can calculate reliably both shading and reflection effects.

For further validation, Sun's trajectories returned by the code were compared with those provided at Ref. 12. The inclination and azimuth angles match the expected ones within 1% in all tested cases. The code can also match very well values of solar insolation from tables for locations where weather is not an important variable, since no weather correction is taken into account in our code. Fig. S2 shows a comparison between the simulated ground insolation on the 15$^{th}$ day of each calendar month - obtained from simulation of a flat horizontal panel of area 1 m$^2$ with 100% efficiency and zero reflectivity – and the one reported by insolation tables obtained at Ref. 27 for each given month by averaging insolation values over the days of that month, for several years. The comparison is shown for two locations: Dubai, where rain and bad weather occur for few days a year, and Boston, where rain and cloudiness are frequent.

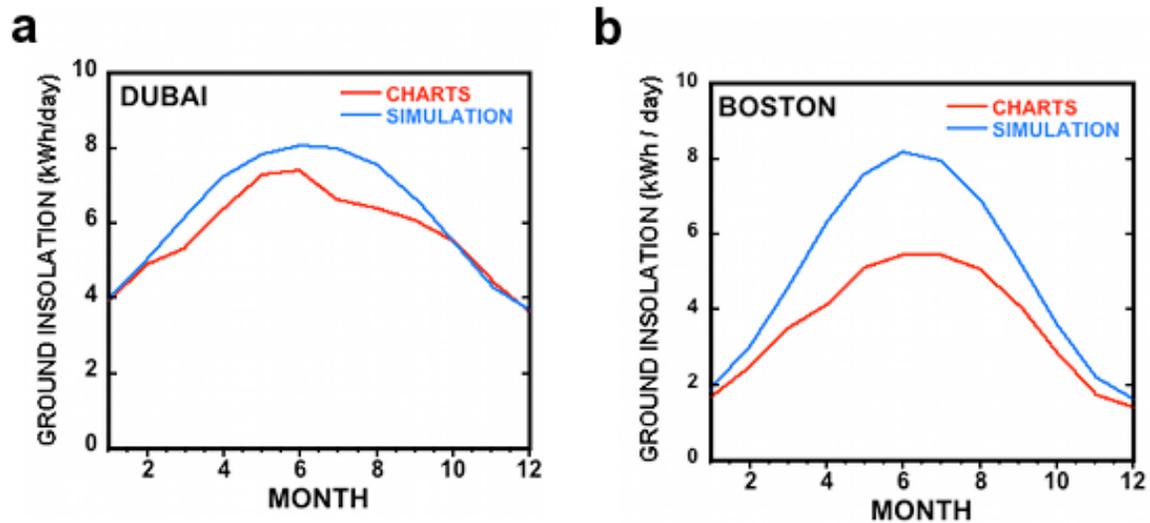

**Fig. S2** (a) Simulated versus measured (Ref. 27) insolation in Dubai, where the weather is clear for most days of the year. Excellent agreement is found between the tables and the simulation, suggesting that the code is reliable for simulation of clear weather conditions. (b) Simulated versus measured (Ref. 27) insolation in Boston, where the simulation overestimates the average measured insolation due to weather corrections not accounted for in the code.

For Dubai (Fig. S2a), the simulation gives a smooth profile that matches extremely well (within 1 %) the insolation chart for 6 months of the year and with deviations within 10 % for the remaining months, likely due to a complex interplay of meteorological conditions beyond the physics captured by our simple AM correction at eq. 3. It must be mentioned, however, that even between different literature sources of insolation charts a discrepancy by 5–10 % is not uncommon.

This suggests that our method is reliable for simulation of clear weather, with excellent predicting capabilities for such conditions. For Boston (Fig. S2b), the simulated insolation exceeds the average insolation for a given month from the charts, due to weather corrections absent in our code. For example, for a rainy day with almost no insolation, there is no contribution to the average insolation reported in the tables, which consequently report lower insolation values. Weather corrections on the other hand don't affect our comparisons between different flat and 3D shapes, and all the results presented in this paper must be understood for a day of clear weather (unless otherwise noted), as mentioned in the paper.

Further validation tests are available upon request.

## 2.2 – DETAILS OF ALL SIMULATIONS

**Simulation of indoor experimental measurements.**

All simulations were performed using a grid of 10,000 points per cell, with a version of the code not implementing AM corrections and using a solar energy flux of 1000 W/m$^2$ to match the emission of the lamp used in the solar simulator. A reflectivity $R$=14 % was assumed based on the epoxy resin coating with refractive index $n = 1.6$ and on known models of the reflectivity of coated Si solar cells (Ref. 28). The code decouples the fate of reflected and transmitted light, and thus in order to reproduce our experiment (where solar cells with AM1.5 efficiency of 10 % – which already includes a reflection loss of 14% the incident energy – were used) the efficiency of the panels was set to $\eta =$ 12 % in the simulations. The cubic 3DPV structure was modeled as a cube of 35 mm side, while the flat panel was modeled as a rectangle of sides 33 mm and 37 mm, as measured experimentally. The position of the Sun was matched to that of the lamp in the experiment by choosing a date and time where the Sun forms a Zenith angle equal to the tilt angle used in the experiment and an azimuth angle of 180° (due South), so to reproduce the rotation around the base edge adopted in the experiment. The power returned at the time when the azimuth angle is 180° was compared to the experimentally measured power. Specific values for each tilt angle were obtained as reported below, and can be checked at Ref. 12:

<u>0 degrees</u>: Latitude 23deg, Longitude -71deg, GMC -5. Jun 31$^{st}$ 11:45AM.

<u>30 degrees</u>: Latitude 43deg, Longitude -71deg, GMC -5. April 26$^{th}$ 11:42AM.

<u>45 degrees</u>: Same location as above. Mar 17$^{th}$ 11:53AM.

<u>60 degrees</u>: Same location as above. Feb 3$^{rd}$ 12:00PM.

The Zenith angle from the simulation was checked to be within 1° of the expected one reported above.

**Simulation of outdoor experimental measurements.**

The same method and cells properties of the indoor simulations were used (see above). The latitude and longitude were set to the values for Cambridge, Massachusetts (Latitude = 42.34°N, Longitude = 71.1°W, GMC -5), and the date was set to be the same as the one when the experiment was performed.

**Concentrator simulation runs.**

Optimization of mixed mirrors and solar cells structures was carried out using the MC algorithm with an inverse-log cooling schedule (see above). Trial moves consisted of translation of 1 coordinate within a cubic box of side length 20 % that of the simulation box, and the solar trajectory for Jun 15$^{th}$ in Boston was used. Initial configurations consisted of random arrangements of 10 triangular cells in a 10x10x10 m$^3$ cubic simulation box. All solar cells were arbitrarily chosen to have efficiency $\eta$=10 % and reflectivity $R$=4 %, and all mirrors had zero efficiency and $R$=100 %. Both the mirrors and the solar cells were considered to be double-sided. The number of mirrors was varied in separate simulations between 0 and 9 (out of a total of 10 panels), and kept constant during each simulation, so to obtain constrained optimization runs with different total mirror and cells areas. After 120,000 simulation steps the energy increased on average by a factor of 10–50 compared to the initial random configuration, and structures with maximal energy generation were extracted and analyzed. Table S1 shows the energy generation and other data for such optimal structures, as discussed in the paper. The simulation reported as "flat panel" shows for comparison data for a flat single-sided horizontal cell covering the base of the simulation box, *i.e.* the solution with maximal energy per solar cell area in the absence of concentration. Note that 3DPV does not normally optimize this figure, but rather optimizes the conversion from a given volume and for a given base area (defined as footprint area of the simulation box, in our case 100 m$^2$), as discussed in the paper.

| SIMULATION NUMBER | 1 | 2 | 3 | 4 | 5 | 6 | 7 | 8 | 9 | FLAT PANEL |
|---|---|---|---|---|---|---|---|---|---|---|
| ENERGY (kWh/day) | 180.7 | 168.7 | 166.8 | 168.3 | 160.8 | 161.2 | 146.3 | 125.0 | 103.3 | 84.3 |
| ENERGY from REFLECTIONS (kWh/day) | 4.8 | 8.0 | 10.0 | 16.2 | 15.5 | 29.3 | 40.5 | 42.0 | 53.5 | 0 |
| MIRRORS AREA (m$^2$) | 0 | 87 | 117 | 127 | 252 | 270 | 404 | 430 | 447 | 0 |
| SOLAR CELLS AREA (m$^2$) | 1003 | 916 | 741 | 736 | 690 | 450 | 405 | 267 | 169.2 | 100 |
| ENERGY / SOLAR CELLS AREA (kWh/m$^2$) | **0.18** | 0.18 | 0.23 | 0.23 | 0.23 | 0.36 | 0.36 | 0.47 | **0.62** | 0.84 |
| ENERGY / FOOTPRINT AREA (kWh/m$^2$) | 1.81 | 1.69 | 1.67 | 1.68 | 1.61 | 1.61 | 1.46 | 1.25 | 1.03 | 0.84 |

**Table S1 | Optimization of structures with mirrors and solar cells.** Concentration of sunlight in a 3DPV structure shows several trends. As the mirror area is increased within a given volume, the energy obtained by a unit area of solar cell increases from 0.18 to 0.62 (see values in bold), and thus by up to a factor of 3.5 compared to a 3D structures without mirrors (first column). For the highest mirror area (simulation #9), an energy per unit PV material of 0.62 kWh/m$^2$ was found, which is almost as high as the flat panel case (0.84 kWh/m$^2$) but with total energy generation higher by 25% compared to the flat case (second row). In this limit, the use of a given amount of PV material is optimal for 3DPV. The presence of mirrors, on the other

hand, decreases the energy per footprint area (*i.e.* the energy density), as seen in the last row of the table and as discussed in the paper.

**Latitude dependent annual energy generation.**

Annual energy generation for 3D structures constituted by cells with 10% efficiency and 4% reflectivity at normal incidence (refraction index *n* = 1.5) was calculated at locations of different latitude between 35° South to 65° North (*i.e.* almost all inhabited land), with an approximate latitude increase of 10° between locations (Table S2). Since little variation was recorded as a function of the longitude (also due to our study not accounting for weather corrections), data of only one location per 10° latitude interval is reported here, and the results are understood valid for locations with a same latitude and arbitrary longitude.

| LOCATION AND LATITUDE (degrees; N=North, S=South) | ENERGY (kWh/m² year) | | | |
|---|---|---|---|---|
| | FLAT HORIZONTAL | OPEN CUBE | FUNNEL | CUBE / FLAT INCREASE FACTOR *Y* |
| Buenos Aires (34 S) Melbourne (37 S) | 184.74 | 475.92 | 491.00 | 2.58 |
| Johannesburg (26 S) | 205.74 | 485.41 | 501.77 | 2.36 |
| Darwin (15 S) | 229.11 | 494.56 | 514.41 | 2.16 |
| Kinhasa (4 S) | 235.39 | 497.63 | 518.52 | 2.11 |
| Bogota (4 N) | 235.74 | 497.59 | 518.59 | 2.11 |
| Bangkok (14 N) Caracas (10 N) | 228.82 | 494.01 | 513.86 | 2.16 |
| Dubai (25 N) Mumbai (19 N) | 210.38 | 487.54 | 504.31 | 2.32 |
| Tokyo (35 N) Mojave Desert (35 N) | 185.60 | 477.98 | 493.25 | 2.58 |
| Boston (42N) Rome (42N) Beijing (40 N) | 165.51 | 464.40 | 477.94 | 2.81 |
| Moscow (55 N) Berlin (52 N) | 123.97 | 416.24 | 426.93 | 3.36 |
| Reykyavik (64 N) | 93.49 | 360.13 | 368.82 | 3.85 |
| Nome (64 N) | | | | |
| Helsinki (60 N) | | | | |

**Table S2 | Annual energy generation of a horizontal flat panel and two 3DPV shapes for different latitudes.** Values of annual energy generation (kWh / m$^2$ year) are shown for different shapes and locations of interest for PV installations. The increase of energy density $Y$ compared to a flat horizontal panel (rightmost column) largely exceeds the one estimated for dual-axis tracking and optimal fixed panel orientation (see paper), remarkably in the absence of any form of dynamic sun tracking for the 3DPV case. Larger increases are found moving away from the Equator towards the Poles, and compensate for the decrease of insolation to yield a much smaller excursion in energy generation at different latitudes compared to the flat horizontal panel case. This reduced seasonal and latitude sensitivity is a built-in feature of 3DPV systems.

Table S2 reports the calculated values shown in Fig. 2a, for an open cube and a funnel (Ref. 5) of 1 m$^2$ base area, and for a flat horizontal panel for comparison. While the open cube is a simple, easy-to-realize 3DPV structure, the funnel has a more advanced design that bears some of the advantages of GA optimized structures,[5] and systematically outperforms most fixed shapes of same volume and base area as confirmed by the data in Table S2. Even for the simple open cube geometry, we observed an increase of the annual energy generation compared to a flat panel by a factor between 2.1–3.8 is shown in Table S2, as reported in the main text.

## 2.3 – POWER BALANCING AND WEATHER EFFECTS

**Power balancing in 3DPV systems.**

The effects deriving from the uneven illumination of solar panels composing a 3DPV system (for example, due to shading by other solar cells) were investigated using a test system consisting of an array of four solar cells (identical to those used in the rest of the work) connected in parallel. Some of the cells in the structure were masked with black electric tape, while one solar cell was illuminated by a natural light source. As an increasing number of cells were covered, the power output decreased progressively, as shown by the I–V curve of the four-cell array (Fig. S3a). We hypothesized this effect may be caused by parasitic dark currents in the masked cells reducing the overall voltage and current, and ultimately reducing the maximum power output of the array.

In order to limit the effect of such parasitic currents, a blocking diode was placed in series with every cell and the array was tested again under the same conditions (Fig. S3b). The losses are seen to almost disappear when this method is used, thus showing that simple blocking diodes can largely mitigate the power unbalances deriving from shaded cells and can be used as an effective tool to reduce electrical losses and optimize the power output of a 3DPV system, at the price of a minimal diode activation voltage. This method was applied for the measurements shown in Fig. 1.

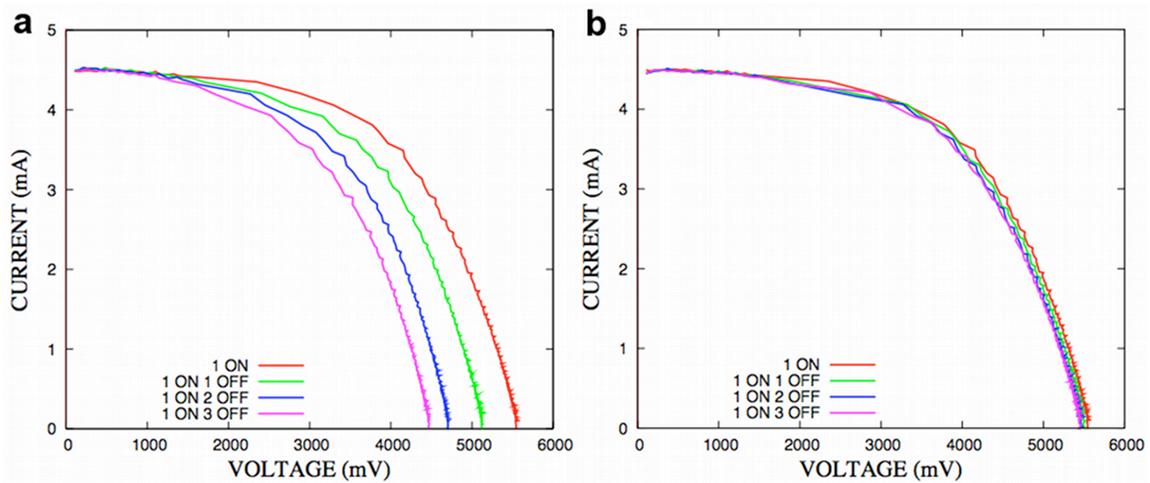

**Fig. S3** (a) Current-voltage curves for an array of four solar cells connected in parallel, for different number of covered and illuminated cells. (b) When blocking diodes connected in series with each cell are added, the power losses are dramatically reduced, with slight residual effects due to the parasitic resistance of the masked solar cells.

**Detail of weather effects on the performance of 3DPV systems.**

The power output of individual 3DPV systems at a given time of the day can be correlated with real-time weather data (for example, from the Weather Bank of the National Climatic Data Center, or the Weather Source and Weather Analytics tool – see Ref. 29 – of the U.S. Department of the Energy). These databases record information regarding precipitations, obscuration, type of distribution of clouds and their elevation, at the location of the experiment. Fig. S4 correlates variations in the power output data with specific events related to cloud conditions, and allows us to extract some trends on the response of 3DPV systems to different meteorological events. We observed, for example, that heavy overcasting at low altitude (below 1500 ft) strongly affects both the 3DPV and the flat panel, while transients with lighter overcasting and substantial diffuse light result in enhancements in the power output of the 3DPV structure over a flat panel compared to a sunny day. Peaks in power generation from 3DPV structures correspond to the presence of few scattered clouds or high altitude overcasting, that likely act as an ideal source of diffuse light. Such effects combine together to yield an increase in the daily energy generation of 3DPV (relative to the flat panel case) even higher in cloudy weather conditions than for clear weather, as discussed in the manuscript.

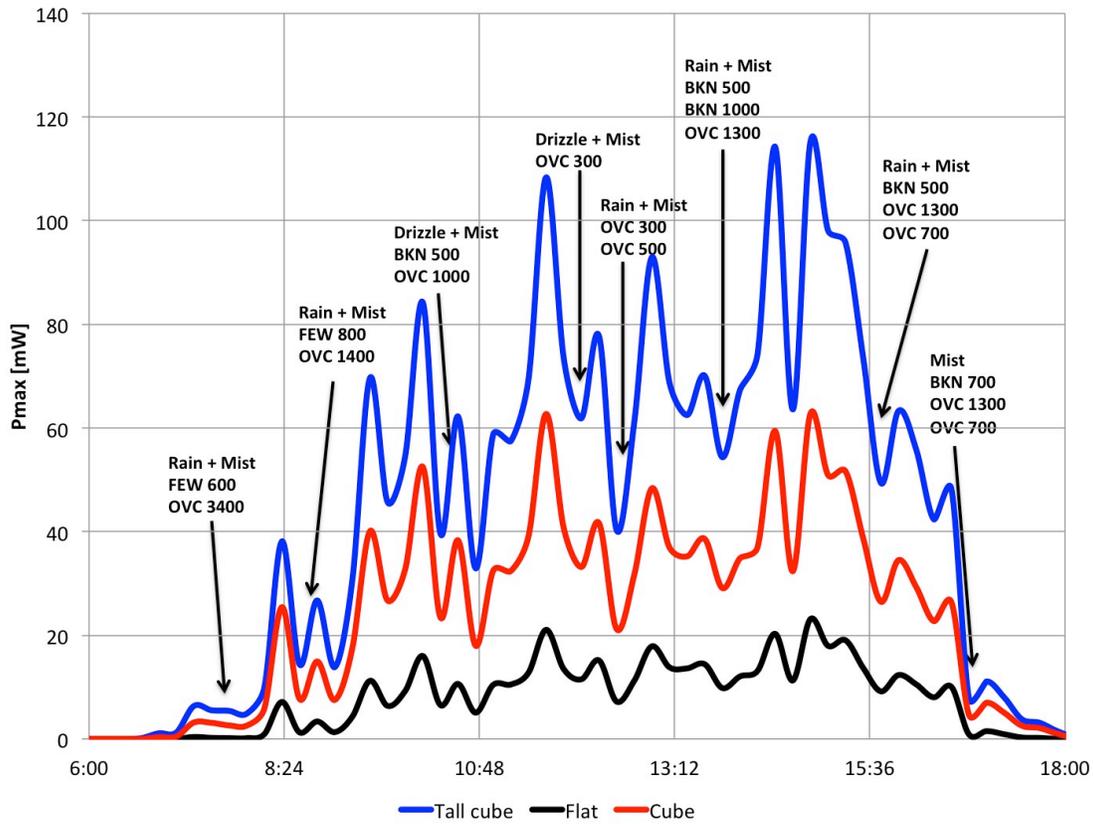

**Fig. S4** Measured power output of 3DPV systems and of a flat panel (for comparison), under overcast, rain, and diffuse insolation conditions. Variations in the power output are associated with weather conditions available from real-time databases. Clouds type and altitude (in feet) are classified according to the Federal Meteorological Handbook (Ref. 30) and using the following International weather codes (Ref. 31): FEW (few clouds), BKN (broken), OVC (overcast).

## 2.4 – EXAMPLES OF APPLICATIONS OF 3DPV

**Sustainable urban environment.**

The absorption of off-peak sunlight and its sensible waste in current rooftop-based technology in urban areas is best understood by comparing the power generated during a day by a tall building (50 m tall parallelepiped with a 10 m side square base area) whose rooftop is coated with 10 % efficient solar panels *vs.* the power generated when all of the building's surface is ideally coated with such solar panels.

Rooftop installation yields the common bell-shaped curve peaked at solar noon, and misses to collect most of the morning and afternoon sunlight that is instead captured by the sides of the structure for the 3D case, with a resulting dramatic increase in the generated power (Fig. S5a). Comparison of the same phenomenon between a day in June and January (Fig. S5b) further shows that 3D sunlight collection and energy generation has reduced seasonal variability (as also discussed in the paper), with a decrease by a factor of 1.65 in the generated energy in going from June to January for the 3D-covered building versus a decrease by a factor of 5.3 for the flat PV case. This trend was found in many other 3D structures we studied. The superior collection of diffused light seems also particularly relevant for application of 3DPV in densely inhabited urban environments where reflected light is conspicuous.

As previously mentioned, the increase in energy density for 3DPV is achieved by using a larger number of solar cells. Simulations for the completely coated building employ 21 times more material than for the case of rooftop coating only. For the winter case (Fig. S5a), an enhancement in energy by a factor of 20 is found, and therefore the solar cells area per unit of generated energy is approximately the same for flat panel design and for 3DPV. This shows how for very tall buildings and for the winter season, the performance of 3DPV can equate that of flat panel design in terms of energy per unit area of active material, with the additional benefit of a significantly higher energy density. For the summer case (Fig. S5b) the active material area per unit of generated energy is higher by a factor of ~3.4 for the 3DPV case compared to the flat panel case.

However, in order to employ all solar cells coating the building and take full advantage of the optimal use of the active material for flat panel design, an area larger than the roof by a factor of 21 would be necessary which is usually not available in residential areas.

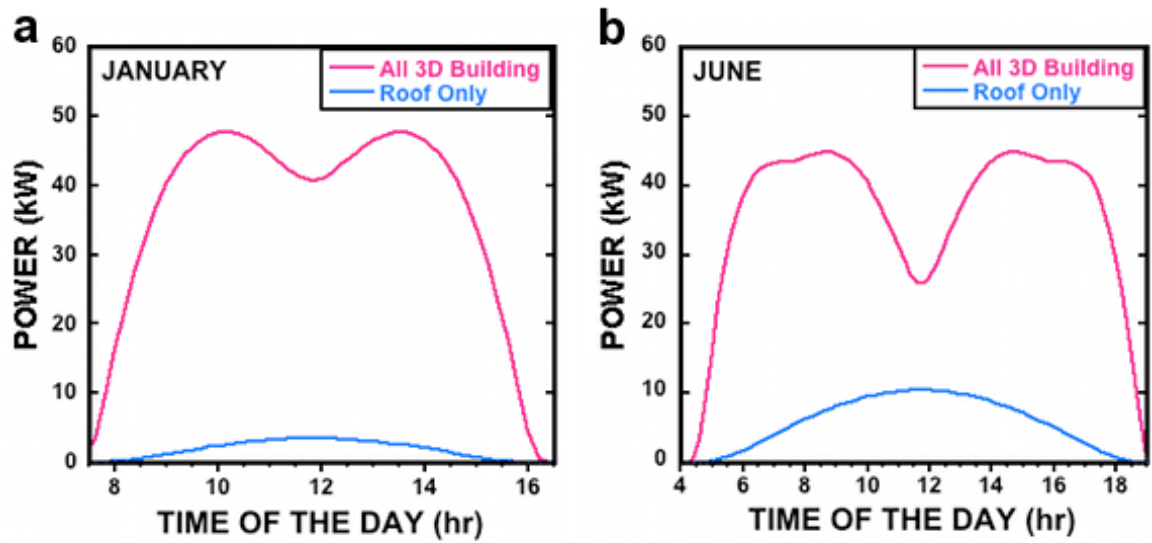

**Fig. S5** (a) Comparison of the power generated by a 50 m tall building completely coated with 10% efficient solar panels (pink curve) versus the same solar cells coating only its rooftop, for a day in January in Boston (Massachusetts, USA). It is apparent how a major loss comes from the bell-shape of the absorption curve for flat PV, missing to exploit most of the morning and afternoon solar energy. The energy generated over the day (integral under each curve) is 16 kWh and 320 kWh for the flat and 3D case, respectively, *i.e.* an increase in energy density by a factor of 20 for the 3D case. This result is relevant to high-energy density generation in sustainable urban environments. (b) Same comparison as in (a), but for a day in June in Boston. The energy generated over the day is 85 kWh and 525 kWh for the flat and 3D case, respectively. Note also how the 3D case is less season-sensitive, given the decrease in generated energy by a factor of 5.3 between June and January for the flat case, and only of 1.65 for the 3D case.

**Design of a 3D e-bike charger.**

As an example of application based on 3DPV currently under study in our lab, we present the design of a 3D charger for electric bicycles (e-bikes). E-bikes are an emerging commodity with a high prospected market demand. The Electric Bikes Worldwide Reports of 2010[32] estimates that 1,000,000 e-bikes will be sold in Europe in 2010, and that sales in the U.S. will reach roughly 300,000 in 2010, doubling the number sold in 2009. We studied a number of shapes for e-bike charging towers based on PV panels arranged in 3D, with a base area of roughly 1 m$^2$ and a height of 4 m for all structures.

Candidate shapes were first designed using CAD software and then obtained as set of triangles that can be studied using our 3DPV code (Fig. S6a,b). The energy generated over a year in Boston was calculated for all candidate shapes, using solar cells with 17 % efficiency and 4 % reflectivity similar to the ones currently in use in our lab to build a prototype. The generated energy ranged between roughly 2–3 MWh/year for optimal shapes orientation, with higher values for designs with "boxy" aspect. Design #4 (Fig. S6a) achieved a maximum simulated energy generation of 3,017 kWh/year, but design #6 – yielding a lower value of 2,525 kWh/year – was selected for the development of a prototype due to weather and wind considerations. These figures are likely in excess by 20–30 % due to weather reducing the annual insolation. The charger can further store power generated during the day in a battery hosted in a compartment placed at the base of the tower (Fig. S6c).

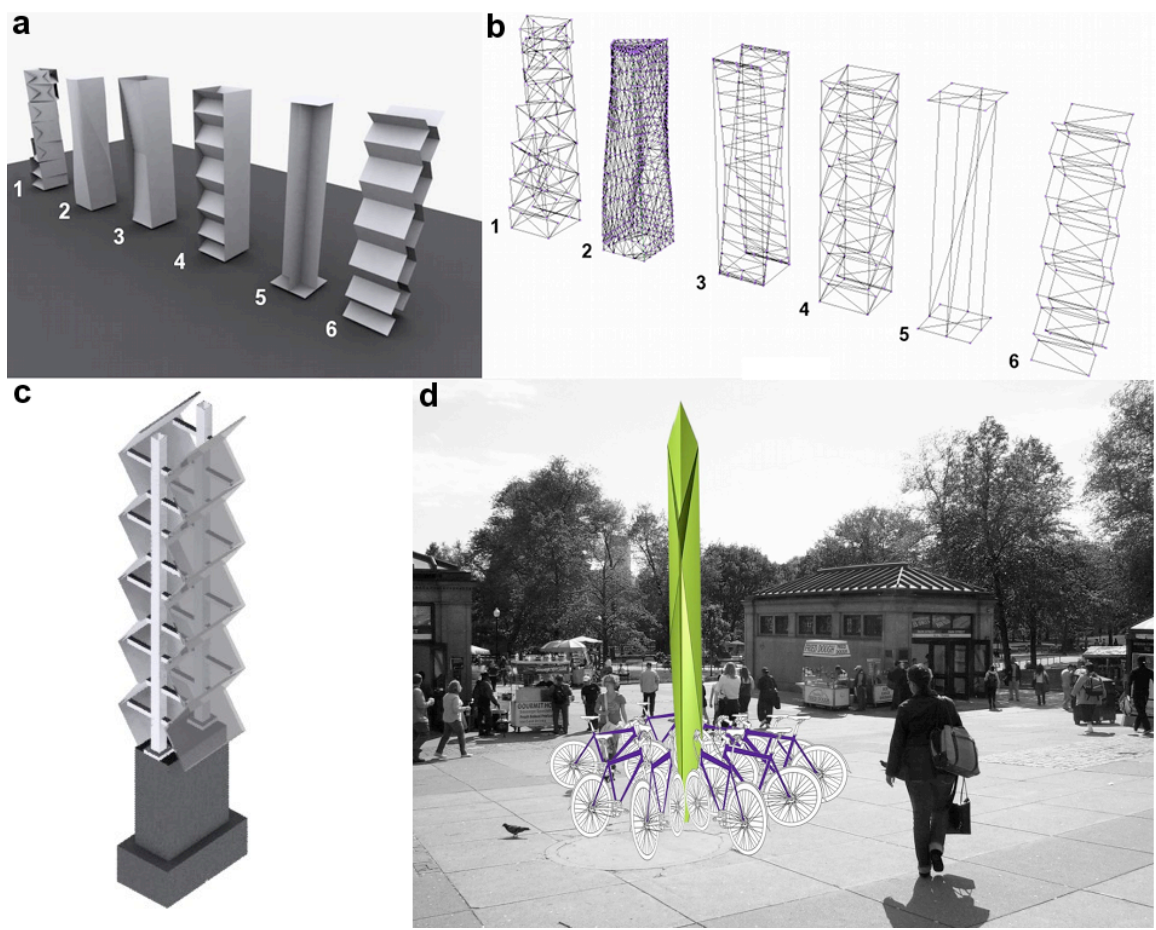

**Fig. S6** (a) Candidate shapes for an e-bike charging tower. (b) Shapes in (a) obtained as set of triangles that can be analyzed with the 3DPV code, with energy generation of 2 – 3 MWh/year. (c) Schematics of a prototype for an e-bike charging station with compact design and easy to integrate in the urban environment. (d) Drawing of an e-bike charging station in the Boston Common and Public Gardens in Boston.

Assuming an ideal charging process and a common energy consumption for e-bikes of 10 – 15 Wh/mile,[32] the tower can charge e-bikes for at least 130,000 miles/year when weather loss are taken into account. Installation of multiple stations in urban areas (Fig. S6d) would facilitate the deployment of e-bike technology, while using charging station of small footprint area, indeed the key feature of 3DPV.

**Supporting References and Notes**